\documentclass[journal]{IEEEtran}

\usepackage{url}
\usepackage{comment}

\ifCLASSINFOpdf
\else
\fi
\usepackage{amssymb, amsmath}

\usepackage{booktabs}

\usepackage{dblfloatfix}
\usepackage[caption=false]{subfig}
\usepackage{tikz}
\usetikzlibrary{positioning, shapes.geometric, arrows.meta, calc, patterns, decorations.pathreplacing}
\newcommand{\archfigscale}{0.70}

\definecolor{trainblue}{RGB}{66, 133, 244}
\definecolor{testorange}{RGB}{234, 170, 0}
\definecolor{attackred}{RGB}{219, 68, 55}
\definecolor{buffergray}{RGB}{180, 180, 180}
\definecolor{excludegray}{RGB}{230, 230, 230}
\definecolor{normblue}{RGB}{200, 220, 255}
\definecolor{darktext}{RGB}{50, 50, 50}

\usepackage{graphicx}
\graphicspath{{figures/}}
\begin{document}
%
\title{Signal Decomposition Reveals Structure in Insider Threat Detection under Sparse Temporal Data}
%
%
%

\author{Hayden~Beadles and Jericho~Cain
\thanks{These authors contributed equally to this work.}%
\thanks{Hayden Beadles is with Georgia Institute of Technology, Atlanta, GA, USA (e-mail: beadles@gatech.edu).}%
\thanks{Jericho Cain is with the Physics Department, Portland Community College, Portland, OR, USA (e-mail: jericho.cain@pcc.edu).}}%

\maketitle

\begin{abstract}
Insider threat detection is difficult not only because malicious behavior is rare, but because it is temporally uneven and embedded within long stretches of inactivity. In enterprise audit data, most windows contain little to no activity, while attacks appear intermittently and vary in duration from brief events to sustained campaigns. Under these conditions, standard reconstruction-based models are dominated by inactive regions and tend to learn baseline behavior rather than the structure of meaningful deviations. We address this by separating activity presence from activity magnitude. Each window is decomposed into a binary mask indicating whether activity occurs and a corresponding value matrix capturing its intensity. A dual-channel autoencoder is trained to reconstruct both components, with the value loss applied only where activity is present, shifting the learning signal toward sparse behavioral structure. Using the CERT r5.2 insider threat datasets as a controlled setting, we examine how anomaly signal manifests under different temporal configurations. Varying window duration and introducing noise reveals consistent patterns: short-duration attacks are detected primarily through presence, longer-duration attacks introduce a measurable magnitude component, and noise suppresses magnitude reliability, forcing the model to revert to presence-dominated detection. The relative contribution of each channel is not fixed, but emerges from the structure of the data. These results suggest a simple interpretation: anomaly signal in sparse temporal data is governed by the interaction between sparsity, temporal extent, and noise. The dual-channel formulation exposes this structure, allowing a single model to adapt across regimes without architectural changes. At the campaign level, where signal is concentrated in a small number of anomalous windows, simple aggregation strategies that emphasize extreme evidence are sufficient to recover extended activity without explicit sequence modeling. Overall, the results indicate that effective insider threat detection depends less on increasing model complexity and more on aligning representation and objective with the structure of sparse temporal data.
\end{abstract}


%
\IEEEpeerreviewmaketitle

\section{Introduction}

Insider threat detection remains a persistent challenge in enterprise security, not because malicious behavior is inherently complex, but because it is rare, intermittent, and embedded within large volumes of normal activity. An insider threat comprises any person with authorized access to an organization's assets (networks, systems, or data) who uses that access, intentionally or unintentionally, to harm that organization. Insider threat attacks are on the rise. IBM reports that 41\% of employees have modified internal company tools without being detected, while industry analyses estimate that insider-related data breaches cost organizations approximately 5M USD annually \cite{ibm_insider_threats, ponemon_insider_risks_2023}.

Enterprise audit logs consist of long stretches of inactivity punctuated by short bursts of user actions across modalities such as logon, file access, email, and web activity. When these logs are aggregated into fixed time windows, the resulting representations are highly sparse, with most entries corresponding to the absence of activity. Under these conditions, anomalous behavior does not typically appear as sustained deviation, but rather as a small number of irregular events embedded within otherwise normal patterns.

This sparsity introduces a fundamental mismatch with reconstruction-based anomaly detection. Autoencoders trained on windowed representations optimize reconstruction error over all entries, and in sparse data this objective is dominated by inactive regions. As a result, much of the model capacity is spent modeling zeros rather than the structure of meaningful behavior. Improvements in reconstruction loss may therefore reflect better modeling of inactivity rather than improved detection of rare events.

Beyond sparsity, insider threat data also exhibits substantial temporal heterogeneity. Attack behavior can vary widely in duration, ranging from short, localized events to extended campaigns unfolding over days or weeks. When anomaly detection is performed on fixed windows, the temporal scale of aggregation directly influences what constitutes a detectable signal. Short windows may capture localized deviations but lack sufficient context, while longer windows provide stability at the cost of diluting sparse activity. As a result, the form of anomaly signal is not uniform across settings, but depends on how behavior is distributed over time.

In this work, we take the view that effective detection in this setting requires aligning the representation and learning objective with the structure of the data. We decompose each window into two components: a binary indicator of activity presence and a corresponding matrix of activity magnitudes. A dual-channel convolutional autoencoder is trained to reconstruct both components, with the value reconstruction conditioned on observed activity. This separation shifts the learning signal away from inactive baselines and toward sparse behavioral structure.

Using the CERT r5.2 insider threat dataset as a controlled environment \cite{Glasser2013}, we examine how anomaly signal manifests under varying temporal configurations. By varying window duration and introducing noise as a perturbation, we observe consistent patterns in how the model allocates importance between presence and magnitude. Short-duration attacks are detected primarily through the presence of activity, while longer-duration attacks introduce a measurable contribution from magnitude. As noise increases, magnitude becomes less reliable and the model reverts to presence-dominated detection. These behaviors are not imposed by the model, but emerge from the structure of the data.

These observations motivate a closer examination of how anomaly signal behaves in sparse temporal data. Rather than assuming a fixed notion of anomaly, we seek to understand how different components of the signal become informative under different conditions. The dual-channel formulation exposes this structure, allowing a single model to adapt across regimes without modification to the architecture. At the campaign level, where malicious behavior is often concentrated in a small number of highly anomalous windows, simple aggregation strategies that emphasize extreme evidence are sufficient to recover extended activity without explicit sequence modeling.

In this sense, the goal of this work is not to optimize benchmark performance, but to characterize how anomaly signal behaves under controlled variations in sparsity, temporal structure, and noise. Specifically, we examine how anomaly signal depends on temporal aggregation, how the relative contribution of presence and magnitude shifts across regimes, and how robust these behaviors are under increasing noise.

The contributions of this work are threefold. First, we identify a structural limitation of reconstruction-based anomaly detection in sparse enterprise telemetry and introduce a presence--magnitude decomposition that addresses it. Second, we characterize how anomaly signal shifts across temporal regimes, showing that the relative importance of presence and magnitude emerges from the data rather than being fixed. Third, we demonstrate that campaign-level detection in this setting follows from aggregation of window-level evidence, without requiring complex temporal models.
\section{Related Work}

Insider threat detection has long been recognized as a difficult problem because malicious behavior is rare, heterogeneous, and often embedded within normal user activity \cite{Salem,Greitzer2013}. More recent surveys note the growing use of deep learning methods in this setting, along with the continued reliance on CERT-style datasets for evaluation \cite{Yuan2020}. Because real enterprise telemetry with high-quality labels is difficult to obtain, the CERT datasets remain a common controlled setting for methodological study. As noted by Glasser and Lindauer \cite{Glasser2013}, synthetic insider threat data is useful for validating detection of well-defined anomalies, but should not be treated as a direct proxy for operational deployment. We adopt that view here, using CERT r5.2 as a controlled environment for probing model behavior rather than as a benchmark for deployable performance.

Prior work on CERT has explored a range of unsupervised approaches, but evaluation practices are inconsistent and often difficult to compare directly. For example, Le and Zincir-Heywood \cite{Le2021} report unsupervised ensemble results using an investigation-budget framework, while Lin et al. \cite{Lin2017} emphasize accuracy, recall, and false positive rate, and Liu et al. \cite{Liu2018} report ROC-oriented performance for deep autoencoder-based detection. In highly imbalanced settings such as insider threat detection, these choices can obscure the practical effect of false positive noise. This motivates our use of PR-AUC as the primary metric, alongside ROC-AUC for completeness, following established arguments that precision--recall analysis is more informative than ROC analysis under severe class imbalance \cite{Davis2006,Saito2015,He2009}.

Our modeling approach is also informed by broader work on anomaly detection and sparse temporal data. Reconstruction-based methods such as autoencoders and variational autoencoders are widely used for anomaly detection because they learn compact representations of normal behavior and detect deviations through reconstruction error \cite{Sakurada2014,an2015variational}. However, these methods typically assume that reconstruction quality is uniformly informative across the input. In sparse enterprise telemetry, that assumption is problematic because inactive regions dominate the representation and can overwhelm the learning signal.

This motivates a representation that treats activity presence and activity magnitude as distinct components. Che et al. \cite{che2016recurrentneuralnetworksmultivariate} show that in multivariate time series with structured missingness, whether a variable is observed can itself be informative and should not be collapsed into imputed values. Mei and Eisner \cite{mei2017neuralhawkesprocessneurally} similarly emphasize an event-driven view of sequential data, where sparse and irregular observations are more naturally modeled as discrete events than as dense continuous signals. Together, these perspectives support treating sparse telemetry not as incomplete dense data, but as event structure in which absence, occurrence, and intensity play different roles.

Our campaign-level aggregation is motivated by the observation that insider threat behavior is often concentrated in a small number of highly anomalous windows rather than distributed uniformly across time. This aligns with the set-based perspective of Zaheer et al. \cite{zaheer2018deepsets}, which formalizes permutation-invariant learning over collections and provides a theoretical basis for aggregation strategies that emphasize a small number of salient elements rather than averaging uniformly across all observations. We use this perspective not to replace window-level detection, but to show how strong window-level scores can be aggregated to recover campaign-level behavior without requiring complex temporal models.

\section{Representation and Model}
\label{sec:representation_and_model}
\subsection{Windowed Behavioral Representation and Sparsity}

We consider the problem of unsupervised insider threat detection from enterprise audit logs. Let $U$ denote a set of users, and let 
\begin{equation}
\mathcal{E}_u = \{(e_i, t_i)\}
\label{eq:raw_logs}
\end{equation}
denote the sequence of events generated by user $u \in U$, where each event belongs to a behavioral modality such as logon, file access, device usage, email, or web activity.

To obtain a regular temporal representation, raw events are aggregated into fixed-length windows of duration $W$ hours. Each window is subdivided into $T$ buckets of width $\Delta t = W/T$. For each user $u$ and window $w$, we construct a feature matrix
\begin{equation}
\mathbf{X}_{u,w} \in \mathbb{R}^{T \times F},
\label{eq:feature_matrix}
\end{equation}
where $F$ is the number of behavioral features and each entry $X_{u,w}(t,f)$ represents the count or summary statistic of feature $f$ observed in bucket $t$ of window $w$. Buckets with no activity are explicitly filled with zeros.

This representation is typically highly sparse. Most time buckets contain no activity across most features, reflecting both the intermittent nature of user behavior and the effects of temporal aggregation. As a result, the majority of entries in $X_{u,w}$ correspond to inactivity. In reconstruction-based models, this leads to a learning objective dominated by inactive regions, where improvements in loss may reflect better modeling of zeros rather than meaningful behavioral structure.

\subsection{Mask--Value Decomposition}

To address this imbalance, we decompose each window into two components:
\begin{equation}
\mathbf{M}_{u,w} \in \{0,1\}^{T \times F}, \quad \mathbf{V}_{u,w} \in \mathbb{R}^{T \times F},
\label{eq:decomposition}
\end{equation}
where
\begin{align}
\mathbf{M}_{u,w}(t,f) &= \mathbb{I}[X_{u,w}(t,f) > 0],\\
\mathbf{V}_{u,w}(t,f) &= X_{u,w}(t,f).
\label{eq:mask_value}
\end{align}

The mask $\mathbf{M}_{u,w}$ captures the \textit{presence} of activity, while $\mathbf{V}_{u,w}$ captures the \textit{magnitude} of that activity. This separation reflects the structure of sparse telemetry: the occurrence of activity is rare and often carries most of the discriminative signal, while magnitude varies only within the subset of active entries.

The two components are stacked along the channel dimension to form a two-channel input tensor
\begin{equation}
\mathbf{X}^{(2)}_{u,w} = [\mathbf{M}_{u,w}, \mathbf{V}_{u,w}] \in \mathbb{R}^{2 \times T \times F}.
\label{eq:two_channel_tensor}
\end{equation}

\subsection{Dual-Channel Autoencoder Architecture}
\begin{figure}[t]
  \centering
  \scalebox{\archfigscale}{
  \begin{tikzpicture}[
      font=\small,
      box/.style={draw, rounded corners, align=center, minimum height=10mm, minimum width=26mm},
      widebox/.style={draw, rounded corners, align=center, minimum height=12mm, minimum width=42mm},
      loss/.style={draw, rounded corners, align=center, minimum height=9mm, minimum width=28mm},
      arrow/.style={-Latex, thick},
      dashedarrow/.style={-Latex, thick, dashed}
  ]

    \node[box] (maskin) {Mask channel\\$\mathbf{M}_{u,w}\in\{0,1\}^{T\times F}$};
    \node[box, right=10mm of maskin] (valin) {Value channel\\$\mathbf{V}_{u,w}\in\mathbb{R}^{T\times F}$\\(z-scored)};

    \node[widebox, below=9mm of $(maskin)!0.5!(valin)$] (stack)
      {Stack channels\\$\mathbf{X}^{(2)}_{u,w}=[\mathbf{M}_{u,w},\mathbf{V}_{u,w}]\in\mathbb{R}^{2\times T\times F}$};

    \node[widebox, below=10mm of stack] (enc) {CNN Encoder\\$f_{\mathrm{enc}}(\cdot)$};
    \node[box, below=10mm of enc] (latent) {Latent code\\$\mathbf{z}_{u,w}\in\mathbb{R}^{d}$};
    \node[widebox, below=10mm of latent] (dec) {CNN Decoder\\$f_{\mathrm{dec}}(\cdot)$};

    \node[box, below left=10mm and 10mm of dec] (maskout) {Mask logits\\$\hat{\mathbf{M}}_{u,w}$};
    \node[box, below right=10mm and 10mm of dec] (valout) {Value recon\\$\hat{\mathbf{V}}_{u,w}$};

    \node[loss, below=10mm of maskout] (bce) {$\mathcal{L}_{\mathrm{mask}}$\\BCE w/ pos\_weight};
    \node[loss, below=10mm of valout] (mse) {$\mathcal{L}_{\mathrm{value}}$\\Masked MSE on $\mathbf{M}=1$};

    \node[loss, right=16mm of latent] (temp) {$\mathcal{L}_{\mathrm{temp}}$\\$\|\mathbf{z}_{u,w+1}-\mathbf{z}_{u,w}\|_2^2$\\(normal pairs)};

    \draw[arrow] (maskin) -- (stack);
    \draw[arrow] (valin) -- (stack);

    \draw[arrow] (stack) -- (enc);
    \draw[arrow] (enc) -- (latent);
    \draw[arrow] (latent) -- (dec);

    \draw[arrow] (dec) -- (maskout);
    \draw[arrow] (dec) -- (valout);

    \draw[arrow] (maskout) -- (bce);
    \draw[arrow] (valout) -- (mse);

    \draw[dashedarrow] (latent) -- (temp);

    \node[align=center, below=7mm of $(bce)!0.5!(mse)$] (total)
      {$\mathcal{L}=\mathcal{L}_{\mathrm{mask}}+\lambda_{\mathrm{value}}\mathcal{L}_{\mathrm{value}}+\lambda_{\mathrm{temp}}\mathcal{L}_{\mathrm{temp}}$};
  \end{tikzpicture}
  }
  \caption{Dual-channel masked autoencoder for window-level UEBA. Each window is decomposed into a binary activity mask and a value matrix. The encoder maps both channels to a shared latent representation, and the decoder reconstructs both channels. Separate losses are applied to the mask and value outputs; an optional temporal consistency term penalizes latent changes between consecutive normal windows.}
  \label{fig:masked_ae_arch}
\end{figure}
We employ a convolutional autoencoder to encode each window into a latent representation shown in Fig. \ref{fig:masked_ae_arch}. The encoder consists of convolutional layers operating over the temporal and feature dimensions, followed by a projection into a latent space:
\begin{equation}
z_{u,w} = f_{\text{enc}}(\mathbf{X}^{(2)}_{u,w}), \quad z_{u,w} \in \mathbb{R}^d.
\label{eq:latent_space_projection}
\end{equation}

The decoder mirrors the encoder and reconstructs both channels:
\begin{equation}
\hat{\mathbf{M}}_{u,w}, \hat{\mathbf{V}}_{u,w} = f_{\text{dec}}(z_{u,w}),
\label{eq:reconstruct}
\end{equation}
where $\hat{\mathbf{M}}_{u,w}$ represents logits for the reconstructed mask and $\hat{V}_{u,w}$ represents reconstructed activity magnitudes.

The value channel $V_{u,w}$ is standardized using statistics computed from training data, while the mask channel remains binary. This allows the model to reason separately about where activity occurs and how much activity occurs, while sharing a common latent representation.

\subsection{Training Objective}

The model is trained exclusively on windows assumed to be normal. The total loss combines three components:
\begin{equation}
\mathcal{L} = \mathcal{L}_{\text{mask}} + \lambda_{\text{value}} \mathcal{L}_{\text{value}} + \lambda_{\text{temp}} \mathcal{L}_{\text{temp}}.
\label{eq:combined_loss}
\end{equation}

\paragraph{Mask Reconstruction Loss.}
The mask loss encourages accurate reconstruction of activity presence and is defined as a weighted binary cross-entropy:
\begin{equation}
\mathcal{L}_{\text{mask}} = \text{BCEWithLogits}(\hat{M}_{u,w}, M_{u,w}; \text{pos\_weight}),
\label{eq:mask_loss}
\end{equation}
where the positive class weight compensates for the imbalance between inactive and active entries. This term dominates the learning signal in sparse settings.

\paragraph{Value Reconstruction Loss.}
The value loss penalizes reconstruction error only where activity is present:
\begin{equation}
\mathcal{L}_{\text{value}} = \frac{1}{|\Omega|} \sum_{(t,f)\in\Omega} \left(\hat{V}_{u,w}(t,f) - V_{u,w}(t,f)\right)^2,
\label{eq:value_loss}
\end{equation}
where $\Omega = \{(t,f) : M_{u,w}(t,f) = 1\}$. This prevents the model from allocating capacity to inactive regions.

\paragraph{Temporal Consistency Regularization.}
To encourage smooth latent trajectories, we optionally apply a temporal regularizer between consecutive normal windows:
\begin{equation}
\mathcal{L}_{\text{temp}} = \mathbb{E}\left[\|z_{u,w+1} - z_{u,w}\|_2^2\right].
\label{eq:temporal_regularizer}
\end{equation}

This term penalizes large changes in latent representations across adjacent windows and is applied only during training on normal data.

\medskip

This formulation separates the learning problem into components aligned with the structure of sparse behavioral data. The mask channel captures the occurrence of activity, while the value channel refines the representation within active regions. As shown in the following sections, the relative importance of these components is not fixed, but varies systematically with the temporal structure of the data.

\subsection{Anomaly Scoring}

To evaluate windows we reconstruct the value and mask matrices for the magnitude $\mathbf{V_{u,w}}$ and presence $\mathbf{M_{u,w}}$, and then calculate the BCE scores for each of them. 

Weights are calculated against each of the components via grid-search, which maximizes the effect of PR-AUC

\paragraph{Combined Anomaly Score.}
\label{sec:combined_anomaly_score}
Each test window produces two raw score vectors: the per-window mask
BCE $\mathcal{L}_{\text{mask}}^{(i)}$ and the per-window active-cell
value MSE $\mathcal{L}_{\text{value}}^{(i)}$. Before combining, each component is
z-standardized across the test set to make the grid search
scale-invariant:
\[
    \widetilde{c}^{(i)}
        = \frac{c^{(i)} - \bar{c}}{\sigma_c + \epsilon},
    \qquad
    c \in \{\mathcal{L}_{\text{mask}},\;
            \mathcal{L}_{\text{value}} \}    \qquad
    \epsilon = 10^{-10}.
\]
The final anomaly score is a weighted sum of these two standardized
components:
\[
    S^{(i)}
        = \alpha_{\text{mask}} \cdot \widetilde{\mathcal{L}}_{\text{mask}}^{(i)}
        + \alpha_{\text{value}} \cdot \widetilde{\mathcal{L}}_{\text{value}}^{(i)}
\]
where $(\alpha_{\text{mask}}, \alpha_{\text{value}}, \beta)$ are
determined by exhaustive grid search maximizing PR-AUC over the
test set.

In addition, we calculate the raw reconstruction PR-AUC without the weights by simply combining $\mathcal{L}_{\text{mask}} + \mathcal{L}_{\text{value}}$. We provide that, along with the ROC-AUC curve (ROC-AUC is computed using the separated, weighted scores). 
\begin{figure*}[t]
\centering
\begin{tikzpicture}[
    x=0.78cm, y=1cm,
    zone/.style={minimum height=1.1cm, anchor=south west, draw=none},
    lbl/.style={font=\footnotesize, text=darktext},
    annot/.style={font=\scriptsize, text=darktext, align=center},
]

\draw[-{Stealth[length=3mm]}, thick, darktext]
    (0, 0) -- (19.5, 0) node[right, font=\small] {time};

\fill[trainblue!30] (0.2, 0.15) rectangle (4.0, 1.25);
\draw[trainblue!70, thick] (0.2, 0.15) rectangle (4.0, 1.25);

\draw[dashed, thick, trainblue!80] (3.05, 0.15) -- (3.05, 1.25);
\node[annot, above] at (1.6, 1.3) {Train ($y{=}0$)};
\node[annot, above] at (3.55, 1.3) {\footnotesize Test};

\fill[excludegray] (4.2, 0.15) rectangle (6.0, 1.25);
\draw[buffergray, thick] (4.2, 0.15) rectangle (6.0, 1.25);
\path[pattern=north east lines, pattern color=buffergray]
    (4.2, 0.15) rectangle (6.0, 1.25);

\fill[attackred!25] (6.2, 0.15) rectangle (10.5, 1.25);
\draw[attackred!80, thick] (6.2, 0.15) rectangle (10.5, 1.25);

\foreach \x in {6.2, 7.3, 8.4, 9.4} {
    \draw[attackred!50, thin] (\x, 0.15) rectangle (\x+0.9, 1.25);
}

\fill[excludegray] (10.7, 0.15) rectangle (12.5, 1.25);
\draw[buffergray, thick] (10.7, 0.15) rectangle (12.5, 1.25);
\path[pattern=north east lines, pattern color=buffergray]
    (10.7, 0.15) rectangle (12.5, 1.25);

\fill[trainblue!30] (12.7, 0.15) rectangle (18.5, 1.25);
\draw[trainblue!70, thick] (12.7, 0.15) rectangle (18.5, 1.25);

\draw[dashed, thick, trainblue!80] (17.3, 0.15) -- (17.3, 1.25);
\node[annot, above] at (15.0, 1.3) {Train ($y{=}0$)};
\node[annot, above] at (17.9, 1.3) {\footnotesize Test};

\node[lbl, below] at (5.1, -0.05) {Buffer};
\node[lbl, below, text=attackred] at (8.35, -0.05)
    {\textbf{Attack Windows}};
\node[lbl, below] at (11.6, -0.05) {Buffer};
\node[lbl, below] at (15.6, -0.05) {Normal};

\draw[-{Stealth}, thick, attackred]
    (8.35, -0.6) -- (8.35, -1.6)
    node[below, annot, text=attackred] {Test set \textbf{only} ($y{=}1$)};

\draw[-{Stealth}, thick, trainblue]
    (1.6, -0.6) -- (1.6, -1.6)
    node[below, annot, text=trainblue]
    {$\mathcal{D}_{\text{train}}$ ($y{=}0$)};

\draw[-{Stealth}, thick, testorange]
    (17.9, -0.6) -- (17.9, -1.6)
    node[below, annot, text=testorange]
    {$\mathcal{D}_{\text{test}}$ ($y{=}0$)};

\draw[{Stealth}-{Stealth}, thick, buffergray]
    (4.2, -0.45) -- (6.0, -0.45);
\node[annot, below] at (5.1, -0.55) {$\Delta_{\text{buf}}$};

\draw[{Stealth}-{Stealth}, thick, buffergray]
    (10.7, -0.45) -- (12.5, -0.45);
\node[annot, below] at (11.6, -0.55) {$\Delta_{\text{buf}}$};

\node[lbl, above] at (6.2, 1.3) {$t_u^{\text{start}}$};
\node[lbl, above] at (10.5, 1.3) {$t_u^{\text{end}}$};

\draw[decorate, decoration={brace, amplitude=5pt, mirror}, thick, trainblue!70]
    (0.2, -0.15) -- (3.05, -0.15)
    node[midway, below=6pt, annot, text=trainblue] {$\lfloor \rho \cdot n_u \rfloor$};

\end{tikzpicture}

\caption{\textbf{Per-user timeline for a malicious user.}
Normal windows (blue) are split chronologically at ratio $\rho = 0.8$:
earlier windows train the autoencoder, later windows test reconstruction of
normal behavior. Attack windows (red) are assigned exclusively to the test set
with label $y{=}1$. Buffer zones (hatched gray) of $\Delta_{\text{buf}}$
hours on each side of the attack are excluded entirely to prevent temporal
leakage.}
\label{fig:timeline}
\end{figure*}
\section{Experimental Framework}
\label{sec:experimental}
\subsection{Dataset}

All experiments are conducted on the CERT r5.2 insider threat datasets, which provide synthetic but behaviorally structured audit logs across multiple modalities, including logon activity, device usage, file access, email, and web traffic for 2000 normal users and 99 malicious users. While synthetic, CERT is widely used in insider threat research due to the difficulty of obtaining real-world labeled data \cite{Glasser2013}.

In this work, CERT is used to define controlled experimental conditions through its scenario structure. Each scenario corresponds to a distinct pattern of insider behavior with different temporal characteristics, particularly in attack duration and activity density. This allows us to evaluate how the model responds to variations in temporal extent and sparsity under consistent data generation assumptions.

CERT comprises a set of data sources \emph{(login, email, http, file, device)} which log user activity over a temporal sequence Eq.~\ref{eq:raw_logs}. Each malicious user $u$ is labeled with an attack window $[t_u^{start}, t_u^{end}]$. 

We train the autoencoder exclusively on normal behavioral windows of the \emph{malicious} users, before and after the attack, with optional added noise injected via normal users, which grow based on the dataset in question. (CERT 4.2 has 1000 users, 70 malicious, CERT 5.2 has 2000 users, and 99 malicious users). 

For each malicious user $u$ with attack interval $[t_u^{\text{start}}, t_u^{\text{end}}]$, we separate the user's behavioral timeline into five zones. 

\begin{itemize}
    \item \textbf{Normal-before}: windows ending before
          $t_u^{\text{start}} - \Delta_{\text{buf}}$.
          Eligible for training and normal-test.
    \item \textbf{Pre-attack buffer}: the interval
          $[t_u^{\text{start}} - \Delta_{\text{buf}},\; t_u^{\text{start}})$.
          Excluded from both sets.
    \item \textbf{Attack interval}: windows overlapping
          $[t_u^{\text{start}}, t_u^{\text{end}})$.
          Labeled $y = 1$ and assigned to the \textbf{test set only}.
    \item \textbf{Post-attack buffer}: the interval
          $(t_u^{\text{end}},\; t_u^{\text{end}} + \Delta_{\text{buf}}]$.
          Excluded from both sets.
    \item \textbf{Normal-after}: windows starting after
          $t_u^{\text{end}} + \Delta_{\text{buf}}$.
          Eligible for training and normal-test.
\end{itemize}

The malicious user events are aligned using the timeline in figure \ref{fig:timeline}, only normal-before and normal-after windows are utilized in training. 

We evaluate the model across multiple temporal resolutions by varying the window size $\Delta_w$ and bucket size $\Delta_b$. Under this construction, normal windows used for training and testing are defined as:

\begin{equation}
\begin{aligned}
\mathcal{N}_u(\text{train} \mid \text{test}) =
\left\{ X_u^{(i)} :
t_i + \Delta_w \leq t_u^{s} - \Delta_{\text{buf}}\right. \\
\text{or} \\
\left. t_i \geq t_u^{e} + \Delta_{\text{buf}}
\right\}
\end{aligned}
\label{eq:normal_set}
\end{equation}

Attack windows are those overlapping the attack interval:
\begin{equation}
\mathcal{A}_u =
\left\{ X_u^{(i)} :
[t_i,\, t_i + \Delta_w) \cap
[t_u^{\text{s}},\, t_u^{\text{e}}] \neq \emptyset
\right\}
\label{eq:anomaly_set}
\end{equation}

Here, $t_i$ denotes the timestamp of a window, $\Delta_{\text{buf}}$ is a buffer period surrounding the attack interval (set to $168\;\text{hrs}$ by default), and $\Delta_w$ defines the temporal extent of each window (e.g., 24, 48, 96 hours).

The final global datasets are constructed as:
\begin{equation}
\mathcal{D}_{\text{train}} =
\bigcup_{u \in \mathcal{U}} \mathcal{N}_u^{\text{train}},
\qquad
\mathcal{D}_{\text{test}} =
\bigcup_{u \in \mathcal{U}}
\left(\mathcal{N}_u^{\text{test}} \cup \mathcal{A}_u \right)
\label{eq:global}
\end{equation}

\subsection{Window Construction}

To obtain a regular temporal representation, we aggregate raw events into fixed-length time windows. Specifically, we partition time into windows of duration $W$ hours, each subdivided into $T$ buckets of width $\Delta t = W / T$. For each user $u$ and window $w$, we construct the feature matrix Eq.~\ref{eq:feature_matrix}. Buckets with no activity are explicitly filled with zeros, resulting in highly sparse representations.

Each window is treated as a single observation for window-level detection. Importantly, the aggregation discards fine-grained ordering of events within each bucket, preserving only coarse temporal structure at resolution $\Delta t$.

\subsection{Noise Injection}

We were motivated by prior work on the role of noise in representation learning, in particular \emph{Generalized Denoising Auto-Encoders as Generative Models} \cite{Bengio2013Denoising}. That work introduces the idea of injecting a corrupting signal, typically modeled as Gaussian noise $\mathcal{N}(\mu, \sigma)$, to improve reconstruction and robustness. In the generative formulation, this is represented as a corruption process $C(\tilde{X} \mid X)$ applied to the input data.

To study robustness under realistic conditions, we introduce noise by incorporating additional normal windows from other users into the training and test sets. This increases the volume of benign activity and simulates the effect of unrelated user behavior present in real enterprise environments.

Noise is controlled as a fraction of additional normal data. Increasing noise reduces the signal-to-noise ratio, particularly for short-duration attacks, and provides a controlled way to examine how the model adapts when anomalous behavior becomes more difficult to distinguish from background activity.

\subsection{Evaluation Metrics}
\begin{figure*}[t]
\centering
\subfloat[\emph{$R_{\mathrm{all}} = \{1,2,3,4\}$, $\Delta_w = 24$, $\Delta_b = 1$ - ROC-AUC Curve}]{%
    \includegraphics[width=0.48\textwidth]{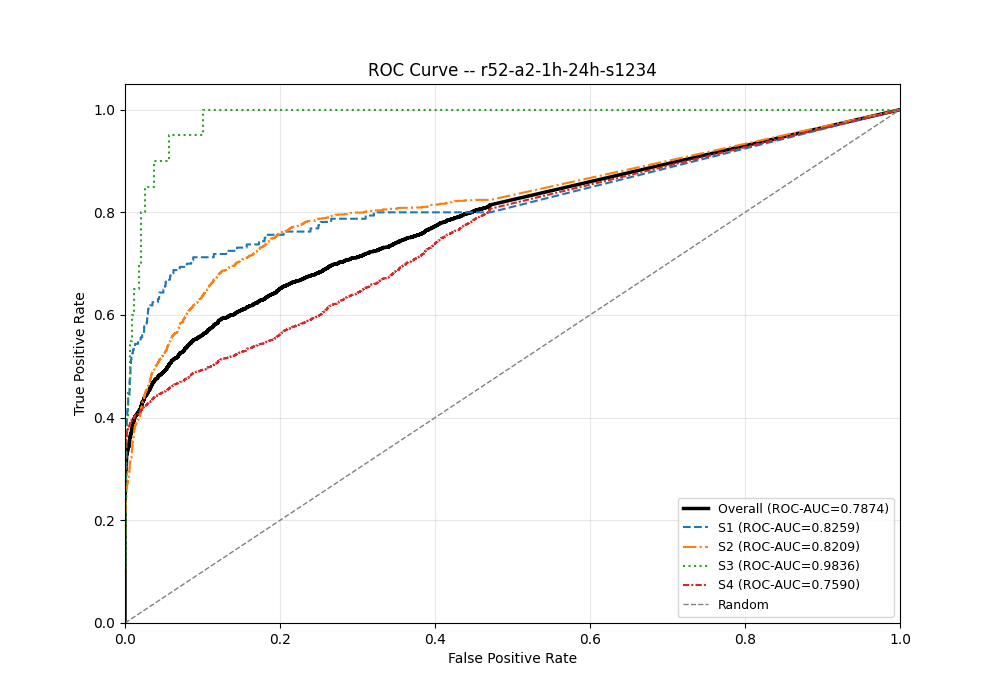}%
    \label{fig:1234-24-1-roc}%
}
\hfill
\subfloat[\emph{$R_{\mathrm{all}} =\{1,2,3,4\}$ $\Delta_w = 24$, $\Delta_b =1$ - PR-AUC Curve}]{%
    \includegraphics[width=0.48\textwidth]{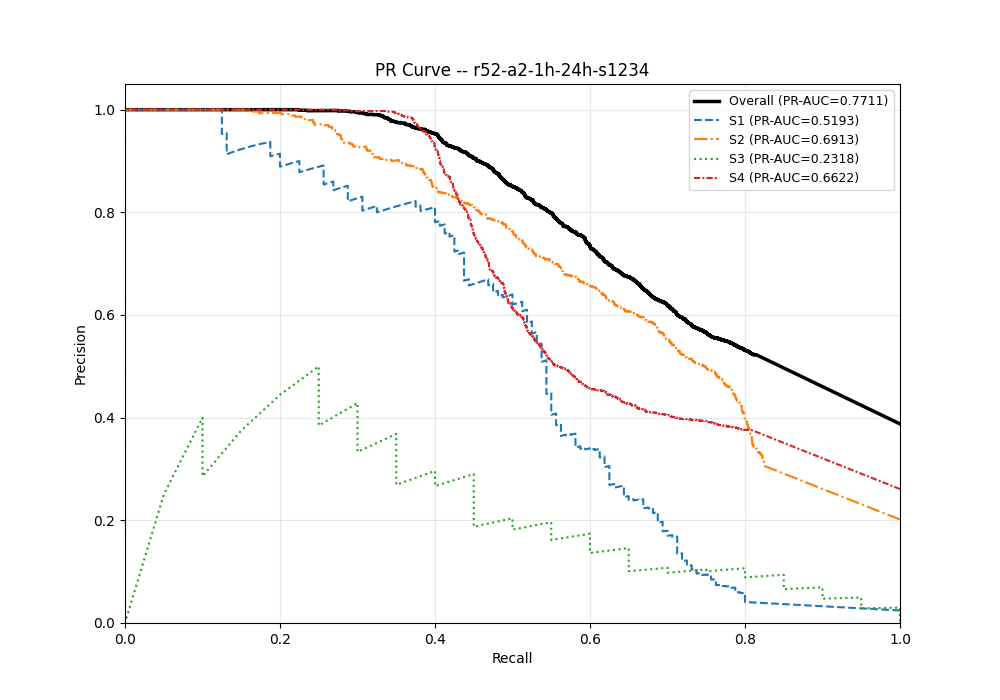}%
    \label{fig:1234-24-1-pr}%
}
\caption{Comparison of ROC-AUC and PR-AUC across all scenarios ($R_{\text{all}} = \{1,2,3,4\}$) for $\Delta_w = 24$ and $\Delta_b = 1$. Scenario 3 achieves a high ROC-AUC of $0.9836$, yet its PR-AUC is only $0.2318$, indicating poor precision despite strong ranking performance. In practice, this corresponds to a high volume of false positives, with approximately $23\%$ of alerts representing true positives.}
\label{fig:all-scenario-summary}
\end{figure*}
\begin{table*}[tb]
\centering
\caption{Experimental Results Across $R_{\mathrm{all}} = \{1,2,3,4\}$}
\label{tab:results}
\renewcommand{\arraystretch}{1.2}
\setlength{\tabcolsep}{4pt}
\footnotesize
\begin{tabular}{cc c ccc cc cccc c}
\hline\hline
$\Delta_b$ & $\Delta_w$ & Noise & PR-AUC & PR-AUC\textsubscript{recon} & ROC-AUC & $\alpha_v$ & $\alpha_{\mathrm{mask}}$ & $\lambda_t$ & $\lambda_v$ & Batch & LR & $d_z$ \\
\hline
1 & 12 & 0 & 0.596 & 0.595 & 0.696 & 0.1 & 0 & 0.0567 & 0.0441 & 32 & 1.66e-04 & 64 \\
1 & 24 & 0 & 0.775 & 0.737 & 0.790 & 0.25 & 0.5 & 0.0687 & 0.0193 & 32 & 1.38e-03 & 96 \\
1 & 32 & 0 & 0.625 & 0.598 & 0.715 & 0.1 & 0.1 & 0.0455 & 0.0216 & 32 & 1.77e-03 & 64 \\
2 & 32 & 0 & 0.656 & 0.585 & 0.720 & 0.5 & 0.75 & 0.0472 & 0.0121 & 32 & 2.02e-03 & 64 \\
1 & 48 & 0 & 0.860 & 0.854 & 0.871 & 0.75 & 0.5 & 0.0281 & 0.0574 & 32 & 1.01e-04 & 96 \\
2 & 48 & 0 & 0.856 & 0.835 & 0.864 & 0.25 & 0.5 & 0.0276 & 0.0409 & 64 & 2.61e-04 & 96 \\
2 & 96 & 0 & 0.887 & 0.852 & 0.903 & 0.5 & 0.75 & 0.0266 & 0.0541 & 32 & 2.46e-03 & 64 \\
1 & 24 & 0.08 & 0.781 & 0.757 & 0.774 & 5 & 5 & 0.0373 & 0.0215 & 64 & 8.98e-04 & 64 \\
\hline\hline
\end{tabular}
\end{table*}
We evaluate detection performance using precision--recall area under the curve (PR-AUC) as the primary metric, with ROC-AUC reported for completeness. In highly imbalanced settings such as insider threat detection, PR-AUC provides a more informative measure of performance, as it reflects the tradeoff between precision and recall in the presence of large numbers of true negatives.

We examined prior research on the CERT dataset using unsupervised approaches, which employs a range of modeling techniques and reporting practices. This makes it challenging to determine which methods are effective. For example, \emph{Anomaly Detection for Insider Threats Using Unsupervised Ensembles} \cite{Le2021} provides a useful reference point, but uses different terminology for standard metrics. In that work, recall is reported as detection rate (DR), and an investigation budget is used as a cumulative distribution function cutoff to determine the number of alerts presented to an analyst.

Prior work on CERT often relies on ROC-AUC or recall as primary evaluation metrics. For example, \cite{Lin2017} reports accuracy, recall, and false positive rate, while \cite{Liu2018} reports ROC-AUC with a similar investigation budget but does not explicitly account for the impact of false positive noise. In highly imbalanced datasets such as CERT, benign activity vastly outnumbers anomalous behavior, so even a low false positive rate can produce a large volume of false alerts that an analyst must process.

For this reason, we emphasize PR-AUC as the primary metric, as it better reflects performance under realistic class imbalance while still reporting ROC-AUC for completeness.

For each window, anomaly scores are computed from reconstruction error components and combined into a final score. Performance is evaluated at the window level by comparing these scores against ground-truth labels. In later sections, we also examine how these scores behave under aggregation for campaign-level detection.

\section{Results}
\label{sec:results}
\begin{figure*}[t]
\centering
\subfloat[\emph{$R_{\mathrm{I}}=\{1,3\}$,  24hr window, 1hr bucket, ROC-AUC Curve}]{%
    \includegraphics[width=0.48\textwidth]{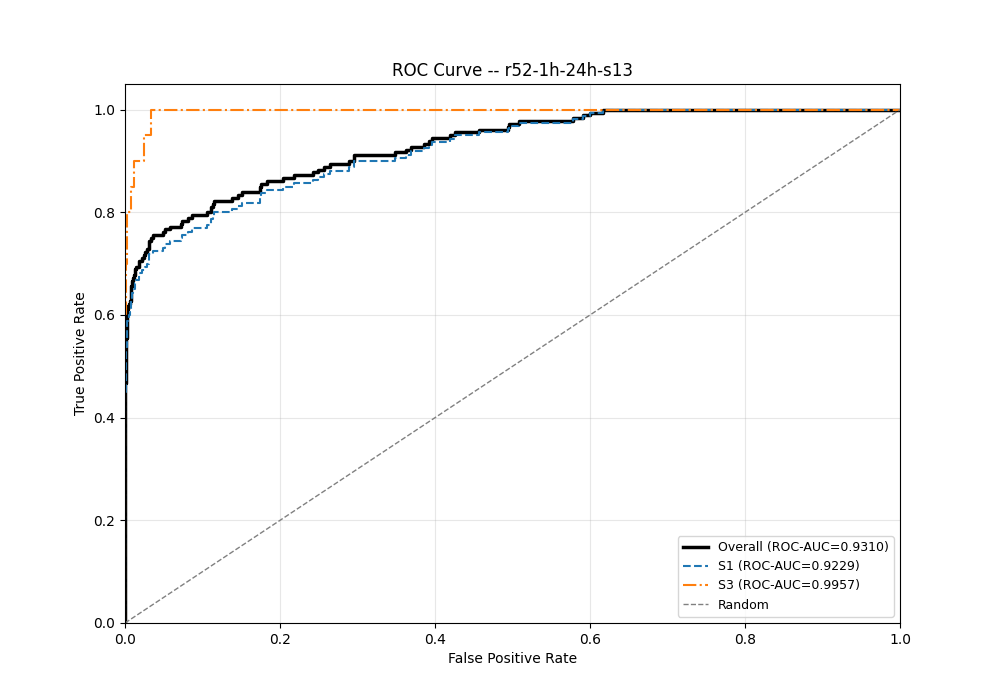}%
    \label{fig:13-24-1-roc}%
}
\hfill
\subfloat[\emph{$R_\mathrm{I}=\{1,3\}$, 24hr window, 1hr bucket, PR-AUC Curve}]{%
    \includegraphics[width=0.48\textwidth]{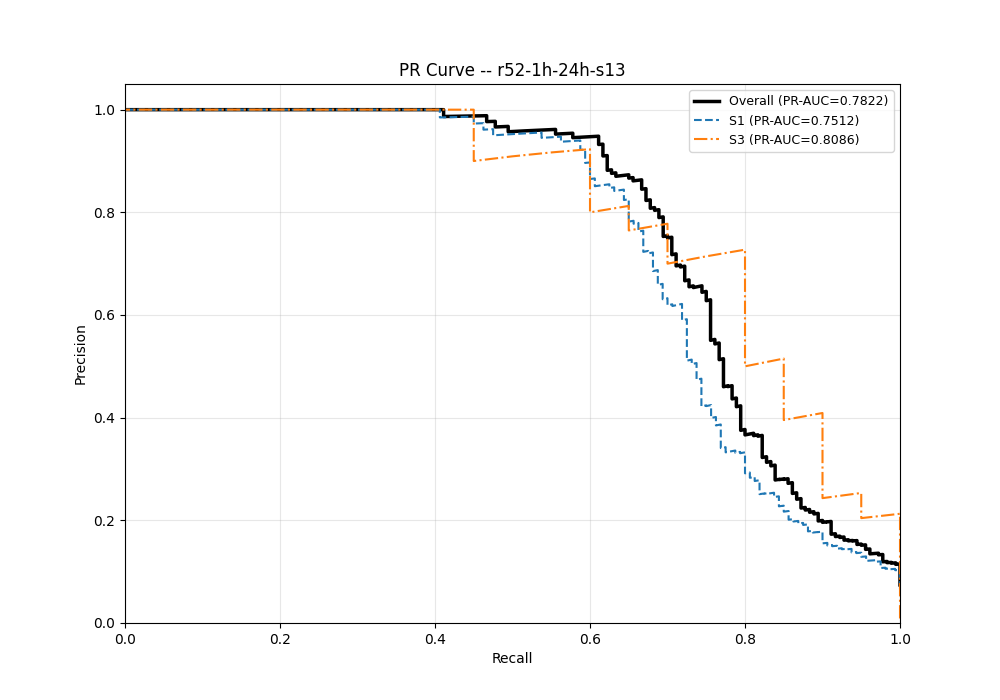}%
    \label{fig:13-24-1-pr}%
}
\caption{Overall best results for $R_\mathrm{I}$. Scenario 3 shows strong ROC-AUC and PR-AUC, in contrast to Fig.~\ref{fig:1234-24-1-pr} where PR-AUC was low despite high ROC-AUC.}
\label{fig:scenario-13-summary}
\end{figure*}
\begin{table*}[tb]
\centering
\caption{Experimental Results for $R_{\mathrm{I}}$}
\label{tab:results_s13}
\renewcommand{\arraystretch}{1.2}
\setlength{\tabcolsep}{4pt}
\footnotesize
\begin{tabular}{cc c ccc cc cccc c}
\hline\hline
$\Delta_b$ & $\Delta_w$ & Noise & PR-AUC & PR-AUC\textsubscript{recon} & ROC-AUC & $\alpha_v$ & $\alpha_{\mathrm{mask}}$ & $\lambda_t$ & $\lambda_v$ & Batch & LR & $d_z$ \\
\hline
1 & 12 & 0 & 0.429 & 0.244 & 0.753 & 0.1 & 5 & 0.0313 & 0.0128 & 32 & 1.46e-03 & 64 \\
1 & 24 & 0 & 0.752 & 0.566 & 0.842 & 0.1 & 3 & 0.0336 & 0.0262 & 32 & 3.61e-03 & 96 \\
2 & 24 & 0 & 0.770 & 0.399 & 0.844 & 0 & 0.1 & 0.0526 & 0.0217 & 32 & 7.99e-04 & 32 \\
1 & 32 & 0 & 0.551 & 0.378 & 0.817 & 0.1 & 0.25 & 0.0157 & 0.0215 & 64 & 2.70e-04 & 96 \\
1 & 12 & 0.08 & 0.353 & 0.160 & 0.737 & 0 & 0.1 & 0.0417 & 0.0170 & 32 & 2.96e-04 & 64 \\
1 & 24 & 0.08 & 0.572 & 0.264 & 0.815 & 0 & 0.1 & 0.0438 & 0.0460 & 32 & 7.04e-04 & 64 \\
\hline\hline
\end{tabular}
\end{table*}
To examine how anomaly signal depends on temporal structure, we evaluate the model across multiple window durations $\Delta_w$ and scenario groupings within the CERT r5.2 dataset. Window size controls the temporal extent over which behavior is aggregated, while the scenarios provide attack patterns with differing durations (Table~\ref{tab:cert52-scenarios}). Together, these factors allow us to probe how the relative contribution of presence and magnitude evolves under different conditions.

We evaluate three experimental settings corresponding to different scenario groupings. First, a model is trained on pooled data from all scenarios (Scenarios 1–4), representing a heterogeneous training regime. We then consider two grouped settings based on attack duration: short-duration scenarios (Scenarios 1 and 3) and long-duration scenarios (Scenarios 2 and 4). This allows us to examine how model behavior changes when training data is aligned with the temporal structure of the underlying attacks.

For each setting, we perform hyperparameter tuning over the search space defined in Table~\ref{tab:search-space}, evaluating performance across window sizes $\Delta_w$ and bucket resolutions $\Delta_b$.

Anomaly scores are computed as a weighted combination of mask reconstruction loss in Eq.~\ref{eq:mask_loss} and value reconstruction error in Eq.~\ref{eq:value_loss}, with weights selected to maximize PR-AUC. Rather than treating these weights as fixed hyperparameters, we analyze how they vary across settings, using them as an indicator of which component of the signal is most informative.

\subsection{Pooled Training Across Heterogeneous Scenarios}

We first examine the effect of training a single model on pooled data from all four scenarios (Scenarios 1--4), while varying the window size $\Delta_w$ and bucket size $\Delta_b$. This setting forces the model to represent attack patterns with substantially different temporal characteristics within a single training regime. As a result, performance reflects a compromise across heterogeneous behaviors rather than alignment to any one attack duration.

Table~\ref{tab:results} summarizes results for this pooled setting, while Figure~\ref{fig:all-scenario-summary} shows representative ROC-AUC and PR-AUC behavior broken out by scenario. Together, these results show that the pooled model tends to perform better on scenarios whose attack durations are more compatible with the chosen temporal resolution, while scenarios with mismatched duration are not well captured.

Two observations are important here. First, PR-AUC reveals substantial variation across scenarios even when ROC-AUC remains uniformly high. This is particularly evident for Scenario 3, where ROC-AUC is strong but PR-AUC is much lower, indicating that the model can rank anomalous windows above normal ones while still producing a large number of false positives in practice. This illustrates why PR-AUC is necessary in this setting, as it more directly captures the tradeoff between detection and false positive noise.

Second, temporal aggregation affects pooled performance in a structured way. As $\Delta_w$ increases, performance generally improves across the pooled setting, but this improvement is not uniform across attack types. The effect of $\Delta_b$ is comparatively smaller: for example, at $\Delta_w = 48$, varying $\Delta_b \in \{1,2\}$ produces similar precision--recall performance. This suggests that temporal extent is more influential than intra-window granularity in the pooled setting.

The pooled model performs noticeably worse at $\Delta_w = 32$. A likely reason is that this window size is poorly aligned with the temporal structure of Scenario 3, combining one full day with only partial information from the next. More generally, this suggests that effective windowing should correspond to meaningful behavioral intervals rather than arbitrary temporal spans.

Across all scenarios, including Scenario 3, the strongest pooled result occurs at $\Delta_w = 24$, where the model achieves PR-AUC $= 0.775$ and ROC-AUC $= 0.790$. At larger window sizes, Scenario 3 is excluded because its attack duration is shorter than the window itself; under those conditions, $\Delta_w = 96$ performs well on Scenarios 1, 2, and 4, reaching PR-AUC $= 0.887$ and ROC-AUC $= 0.903$. This contrast further emphasizes that no single temporal scale is optimal across all attack types.

In this pooled regime, the model also relies on both the presence and magnitude channels. As shown in Table~\ref{tab:results}, $\alpha_v$ and $\alpha_{\text{mask}}$ both contribute to the final anomaly score, with their relative importance shifting across $\Delta_w$ and $\Delta_b$. This is important because it shows that, under heterogeneous training conditions, the model does not settle into a purely presence-driven or purely magnitude-driven solution, but instead balances the two channels depending on temporal configuration.

Overall, Scenario 3 remains the clearest indication of the limitations of pooled training. Its short attack duration makes it difficult to recover when the model is trained across scenarios with much longer temporal structure, highlighting the importance of matching temporal resolution to attack duration. This motivates the regime-based analysis that follows. Rather than training across all attack types simultaneously, we next separate short-duration and long-duration scenarios in order to examine how the relative contribution of presence and magnitude changes when temporal structure is more consistent within the training set.

\subsection{Regime I: Short-Duration Attacks}

We next restrict attention to short-duration scenarios $R_{\mathrm{I}} = \{1,3\}$. These scenarios have comparable attack durations, with Scenario 3 concentrated around $\sim$1.5 days and Scenario 1 spanning shorter to moderately extended intervals. Grouping them allows us to examine whether aligning temporal structure improves the model’s ability to recover short-duration attacks that were poorly captured under pooled training.

Table~\ref{tab:results_s13} summarizes results for this regime, and Figure~\ref{fig:scenario-13-summary} shows the corresponding ROC-AUC and PR-AUC curves. Performance is maximized at $\Delta_w = 24$ with $\Delta_b = 2$, yielding $\text{PR-AUC} = 0.770$ and $\text{ROC-AUC} = 0.844$. As in the pooled setting, $\Delta_w$ has a stronger effect than $\Delta_b$, although increasing $\Delta_b$ provides a modest improvement in this regime.

A key observation is that the model relies almost entirely on the mask channel. Across all configurations in Table~\ref{tab:results_s13}, the hyperparameter search drives $\alpha_v \approx 0$, placing nearly all weight on $\alpha_{\text{mask}}$. This indicates that, for short-duration attacks, the presence of activity carries the discriminative signal, while magnitude contributes little and can act as noise.

This behavior highlights a limitation of single-channel reconstruction approaches. When presence and magnitude are combined, the magnitude component dilutes the more informative presence signal. The dual-channel formulation avoids this by allowing the model to suppress the value component entirely. Consistent with this, the reconstruction-only score (PR-AUC\textsubscript{recon}) is systematically lower than the weighted dual-channel score across all configurations.

The effect of aligning temporal structure is most clearly seen in Scenario 3. Relative to the pooled model (Figure~\ref{fig:1234-24-1-pr}), both ROC-AUC and PR-AUC improve substantially in this regime $(0.9957, 0.8086)$, indicating that separating short-duration scenarios restores precision that was previously obscured by heterogeneous training.
\begin{figure*}[t]
\centering
\subfloat[\emph{$R_{\mathrm{II}}=\{2,4\}$,  96hr window, 1hr bucket, ROC-AUC Curve}]{%
    \includegraphics[width=0.48\textwidth]{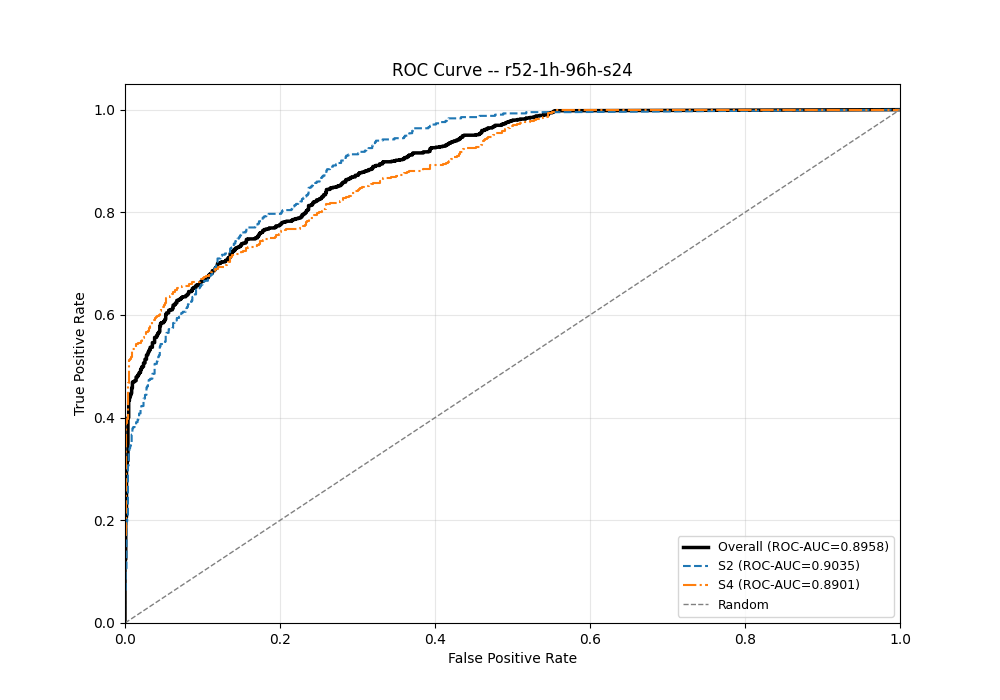}%
    \label{fig:24-96-1-roc}%
}
\hfill
\subfloat[\emph{$R_{\mathrm{II}}=\{2,4\}$, 96hr window, 1hr bucket, PR-AUC Curve}]{%
    \includegraphics[width=0.48\textwidth]{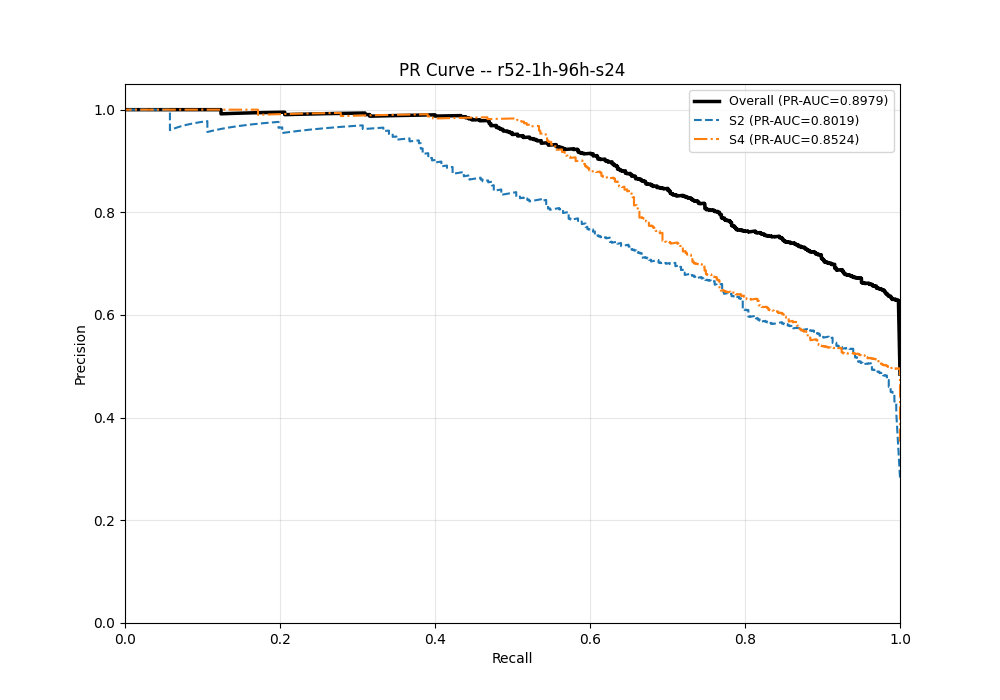}%
    \label{fig:24-96-1-pr}%
}
\caption{Overall best results for $R_{\mathrm{II}}$.}
\end{figure*}
\begin{table*}[!t]
\centering
\caption{Experimental Results for $R_{\mathrm{II}}=\{2,4\}$ Note the model performs best at $\Delta_w = 96$, $\Delta_b$ plays less of an effect. $\alpha_v$ plays more of a role here than in $S_2$, compare to table \ref{tab:results_s13}}
\label{tab:results_s24}
\renewcommand{\arraystretch}{1.2}
\setlength{\tabcolsep}{4pt}
\footnotesize
\begin{tabular}{cc c ccc cc cccc c}
\hline\hline
$\Delta_b$ & $\Delta_w$ & Noise & PR-AUC & PR-AUC\textsubscript{recon} & ROC-AUC & $\alpha_v$ & $\alpha_{\mathrm{mask}}$ & $\lambda_t$ & $\lambda_v$ & Batch & LR & $d_z$ \\
\hline
1 & 12 & 0    & 0.703 & 0.700 & 0.722 & 0.1  & 0    & 0.0224 & 0.0608 & 64 & 3.63e-03 & 64 \\
1 & 24 & 0    & 0.829 & 0.818 & 0.808 & 0.75 & 1    & 0.0352 & 0.0593 & 64 & 1.91e-04 & 96 \\
2 & 24 & 0    & 0.830 & 0.812 & 0.806 & 0.25 & 1    & 0.0544 & 0.0284 & 32 & 6.00e-04 & 96 \\
1 & 48 & 0    & 0.878 & 0.869 & 0.870 & 1    & 0.75 & 0.0557 & 0.0285 & 64 & 7.35e-04 & 96 \\
2 & 48 & 0    & 0.880 & 0.867 & 0.867 & 0.25 & 0.5  & 0.0467 & 0.0558 & 32 & 1.39e-03 & 96 \\
1 & 96 & 0    & 0.906 & 0.896 & 0.902 & 0.1  & 0.25 & 0.0202 & 0.0258 & 32 & 3.63e-04 & 96 \\
2 & 96 & 0    & 0.903 & 0.878 & 0.901 & 0.25 & 0.5  & 0.0648 & 0.0539 & 64 & 6.34e-04 & 96 \\
1 & 12 & 0.08 & 0.709 & 0.675 & 0.711 & 40   & 0    & 0.0181 & 0.0247 & 32 & 1.19e-03 & 64 \\
1 & 24 & 0.08 & 0.823 & 0.792 & 0.802 & 0.25 & 0.75 & 0.0541 & 0.0102 & 32 & 1.27e-03 & 96 \\
1 & 24 & 0.20 & 0.832 & 0.767 & 0.809 & 0    & 0.1  & 0.0426 & 0.0329 & 64 & 1.38e-04 & 64 \\
\hline\hline
\end{tabular}
\end{table*}
\subsection{Regime II: Long-Duration Attacks}

We now consider longer-duration scenarios $R_{\mathrm{II}} = \{2,4\}$. These scenarios exhibit sustained activity over extended periods, requiring the model to capture behavior across a broader temporal horizon. As a result, performance improves as $\Delta_w$ increases, reflecting the benefit of additional temporal context in capturing extended attack structure.

Table~\ref{tab:results_s24} summarizes results for this regime. Unlike $R_{\mathrm{I}}$, where $\alpha_v \approx 0$, the model consistently assigns non-zero weight to both channels. For example, at $\Delta_w = 96$ and $\Delta_b = 1$, the grid search selects $\alpha_v = 0.1$ and $\alpha_{\text{mask}} = 0.25$. At smaller window sizes ($\Delta_w \in \{24, 48\}$), the interaction between the two channels becomes more pronounced, although $\alpha_{\text{mask}}$ remains dominant.

This reflects a fundamental difference in signal structure. Longer-duration attacks produce sustained deviations across multiple windows, introducing redundancy that stabilizes the anomaly signal. In this setting, magnitude becomes informative: it captures how behavior evolves within active regions, complementing the structural signal provided by presence. The mask channel continues to dominate, but the value channel refines the signal rather than acting as noise.

Consistent with this, the model achieves stronger overall performance in $R_{\mathrm{II}}$ than in $R_{\mathrm{I}}$, particularly at larger $\Delta_w$, where extended temporal context aligns with the underlying attack duration.

\subsection{Noise Perturbation}

Finally, we examine the effect of noise, as reflected in the noisy configurations of Tables~\ref{tab:results_s13} and \ref{tab:results_s24}.

The impact of noise differs substantially across regimes. In the pooled setting, $R_{\text{all}}=\{1,2,3,4\}$, introducing noise has minimal effect on PR-AUC (e.g., $0.775 \rightarrow 0.781$), indicating that aggregate performance is largely insensitive to moderate perturbation.

In contrast, short-duration scenarios $R_{\mathrm{I}}=\{1,3\}$ are highly sensitive to noise. At $\Delta_w = 24$, adding noise ($0.08$) reduces PR-AUC from $0.752$ to $0.572$, a substantial decline. The reconstruction-only scores degrade even more sharply, indicating that magnitude-based signal is particularly vulnerable in this regime. This reflects the limited redundancy of short-duration attacks, where anomalous behavior is concentrated in a small number of windows and is easily obscured by background activity.

Long-duration scenarios $R_{\mathrm{II}}=\{2,4\}$ exhibit markedly greater robustness. Across noise levels, PR-AUC remains stable ($0.829 \rightarrow 0.823 \rightarrow 0.832$ for noise $0.00, 0.08, 0.20$), while reconstruction-only performance degrades more noticeably ($0.818 \rightarrow 0.792 \rightarrow 0.762$). This indicates that although magnitude becomes less reliable under noise, the overall signal remains recoverable due to redundancy across windows.

Across regimes, increasing noise consistently reduces the contribution of the value channel. As noise increases, the model shifts weight toward the mask channel, effectively reverting to presence-dominated detection. This reflects the relative stability of the two signals: presence is sparse and robust to background variation, while magnitude is more easily perturbed.

\medskip
Taken together, these results reveal consistent patterns in how anomaly signal manifests across different conditions, particularly as sparsity, temporal extent, and noise are varied. These patterns motivate a closer examination of the structure of anomaly signal, which we analyze in the following section.

\section{Interpretation of Results}
\label{sec:interpretation}
The results in Section~\ref{sec:results} show that anomaly signal in enterprise telemetry is not uniform, but depends on how behavior is distributed across time. In sparse settings with short-duration attacks, anomalous behavior appears as isolated deviations within otherwise inactive windows. Under these conditions, the presence of activity is itself unusual, and detection is driven almost entirely by the mask channel. As attack duration increases, activity becomes more sustained, allowing magnitude to contribute alongside presence. Under noise, magnitude becomes unreliable and the signal collapses back toward presence.

This behavior is directly reflected in the learned weighting between channels. The relative importance of presence and magnitude is not fixed, but emerges from the data. In short-duration scenarios, the value component is consistently suppressed, with $\alpha_v \rightarrow 0$. In longer-duration scenarios, $\alpha_v$ remains non-zero and increases with window size, reflecting the contribution of sustained activity. Under noise, $\alpha_v$ is again reduced as magnitude loses reliability.

The model does not impose this structure. The dual-channel formulation exposes both components of the signal and allows optimization to select between them. In this sense, the weighting parameters act less as tunable design choices and more as indicators of which aspects of the data are informative under a given regime. The model is not combining signals so much as revealing how the signal is composed.

This perspective clarifies why standard reconstruction objectives struggle in sparse data. When inactivity dominates, reconstruction error is driven by baseline behavior rather than meaningful deviations. Separating presence from magnitude isolates where activity occurs from how it behaves, allowing the model to focus capacity on the informative regions of the input.

The same structure also limits the role of temporal modeling. While sequential models can capture long-range dependencies, much of the relevant signal in this setting is localized. Detection depends less on modeling full trajectories and more on identifying and amplifying a small number of high-signal windows.

This concentration of signal has a direct consequence: anomalous behavior is not distributed uniformly over time, but concentrated in a small subset of windows. This observation motivates treating detection at a higher level not as a sequence modeling problem, but as one of aggregating sparse, high-confidence evidence across time.

\section{Campaign Detection}
\label{sec:campaign_detection}
\begin{table}[ht]
\centering
\caption{Campaign-level detection performance using top-$k$ aggregation over six-day sequences (24-hour windows).}
\label{tab:campaign_overall}
\begin{tabular}{lcc}
\toprule
Aggregation Method & PR-AUC & ROC-AUC \\
\midrule
Top-3 Window Scores & 0.7675 & 0.9134 \\
Top-2 Window Scores & \textbf{0.8357} & \textbf{0.9388} \\
\bottomrule
\end{tabular}
\end{table}
While the primary objective of this work is high-precision window-level detection, insider threat behavior typically unfolds over multiple days. In practice, analysts are less interested in isolated anomalous windows than in identifying users engaged in extended attack campaigns. We therefore evaluate whether window-level anomaly scores can be aggregated to support campaign-level detection.

To construct campaign-level examples, we group consecutive windows into fixed-length sequences of $L=6$ days using a sliding window with stride 3. Each sequence is labeled as malicious if it overlaps with at least one anomalous window. Campaign-level anomaly scores are then computed by aggregating the window-level mask reconstruction scores within each sequence.

This formulation reflects the empirical structure observed in Section~\ref{sec:results}, where anomalous behavior is concentrated in a small number of highly informative windows. As a result, aggregation should emphasize extreme evidence rather than uniform averaging.

We therefore consider a simple top-$k$ aggregation strategy, in which the $k$ highest-scoring windows within each sequence are used to form a campaign-level score. This approach captures the intuition that only a small subset of windows contains strong evidence of malicious activity.

As shown in Table~\ref{tab:campaign_overall}, top-$k$ aggregation consistently outperforms broader pooling strategies. In particular, selecting the top 2 windows yields the highest PR-AUC and ROC-AUC, indicating that a small number of high-confidence detections is sufficient to recover extended attack behavior.
\begin{table}[tb]
\centering
\caption{Scenario-specific campaign-level detection performance using top-$k$ aggregation over six-day sequences (24-hour windows).}
\label{tab:campaign_scenarios}
\begin{tabular}{lcccc}
\toprule
 & \multicolumn{2}{c}{Scenario 1} & \multicolumn{2}{c}{Scenario 3} \\
Aggregation Method & PR-AUC & ROC-AUC & PR-AUC & ROC-AUC \\
\midrule
Top-3 Window Scores & 0.7460 & 0.9062 & 0.6179 & 0.9731 \\
Top-2 Window Scores & \textbf{0.8156} & \textbf{0.9327} & \textbf{0.7942} & \textbf{0.9889} \\
\bottomrule
\end{tabular}
\end{table}
Scenario-level results in Table~\ref{tab:campaign_scenarios} show similar improvements across both short-duration scenarios, with top-2 aggregation producing substantial gains in PR-AUC. Under the any-overlap labeling scheme, the expected PR-AUC of a random ranking is approximately equal to the positive class prevalence ($\sim 0.13$), indicating that the observed performance represents a significant improvement over chance despite strong label sparsity.

These results show that campaign-level detection follows directly from the structure of window-level signal. Rather than modeling long-range temporal dependencies explicitly, effective detection can be achieved by aggregating a small number of high-confidence windows. This suggests a natural separation between window-level detection, which identifies anomalous behavior at fine temporal resolution, and campaign-level aggregation, which determines how these signals are combined over time.

\section{Conclusion}
\label{sec:conclusion}

This work presents a unified view of anomaly detection in sparse temporal data, combining a dual-channel representation with a systematic analysis of how anomaly signal behaves under different conditions. By decomposing activity into presence and magnitude, the model aligns the learning objective with the structure of enterprise telemetry and avoids the dominance of inactive regions in reconstruction-based methods.

The results show that anomaly signal in sparse temporal data is structured rather than fixed, with its form determined by how behavior is distributed across time and under noise. Short-duration attacks are detected primarily through presence, longer-duration attacks introduce a meaningful magnitude component, and noise reduces the reliability of magnitude, causing the model to revert to presence-based detection. These behaviors emerge consistently across scenarios and window configurations.

From an operational perspective, the findings suggest a separation between window-level detection and campaign-level aggregation. High-quality window-level scores, when aligned with sparse behavioral structure, can be combined using simple aggregation strategies to recover extended attack behavior without requiring complex temporal models.

These conclusions extend beyond the CERT dataset. Real enterprise telemetry is typically noisier, more heterogeneous, and less structured, but the underlying characteristics observed here -- sparsity, temporal heterogeneity, and localized anomaly signal -- are common to user activity in operational environments. This suggests that effective detection depends less on matching specific patterns and more on aligning representations with these structural properties.

At the same time, several limitations remain. The analysis relies on synthetic data and assumes access to clearly defined attack intervals. Real-world deployments must contend with incomplete labels, evolving user behavior, and domain-specific feature design. While these challenges are not addressed directly, the framework presented here provides a basis for adapting to more complex and less controlled settings.

Future work will extend this analysis to larger and more realistic datasets, including proprietary enterprise telemetry, and will evaluate aggregation strategies across all scenarios using the improved model. Additional directions include exploring alternative decompositions of activity, incorporating contextual features, and studying how representation choices affect robustness under evolving behavior.

Overall, the results indicate that effective insider threat detection depends not only on model design, but on understanding how anomaly signal is structured in sparse temporal data. By focusing on this structure, relatively simple models can adapt across a range of conditions and provide a foundation for scalable detection systems.

\section*{Acknowledgments}

The authors acknowledge the CERT Division of the Software Engineering Institute at Carnegie Mellon University for providing the CERT Insider Threat dataset used in this research. The views expressed in this paper are those of the authors and do not necessarily reflect the views of Carnegie Mellon University or the CERT Division.



\ifCLASSOPTIONcaptionsoff
  \newpage
\fi

\bibliographystyle{iopart-num}
\bibliography{references}

@InBook{Salem,
  author    = {Salem, Malek Ben and Hershkop, Shlomo and Stolfo, Salvatore J.},
  pages     = {69--90},
  year = {2008},
  publisher = {Springer US},
  title     = {A Survey of Insider Attack Detection Research},
  isbn      = {9780387773223},
  booktitle = {Insider Attack and Cyber Security},
  doi       = {10.1007/978-0-387-77322-3_5},
  issn      = {1568-2633},
}

@techreport{ponemon_insider_risks_2023,
  author       = {{Ponemon Institute}},
  title        = {2023 Cost of Insider Risks: Global Report},
  year         = {2023},
  institution  = {Ponemon Institute},
  note         = {Sponsored by DTEX Systems},
  url          = {https://ponemonsullivanreport.com/2023/10/cost-of-insider-risks-global-report-2023/}
}

@misc{ibm_insider_threats,
  author       = {{IBM}},
  title        = {What are Insider Threats?},
  year         = {2025},
  url          = {https://www.ibm.com/think/topics/insider-threats},
  note         = {Accessed: 2026-03-23}
}

@InProceedings{Sakurada2014,
  author     = {Sakurada, Mayu and Yairi, Takehisa},
  booktitle  = {Proceedings of the MLSDA 2014 2nd Workshop on Machine Learning for Sensory Data Analysis},
  title      = {Anomaly Detection Using Autoencoders with Nonlinear Dimensionality Reduction},
  year       = {2014},
  month      = dec,
  pages      = {4--11},
  publisher  = {ACM},
  series     = {MLSDA’14},
  collection = {MLSDA’14},
  doi        = {10.1145/2689746.2689747},
}

@techreport{an2015variational,
  author       = {An, Jinwon and Cho, Sungzoon},
  title        = {Variational Autoencoder based Anomaly Detection using Reconstruction Probability},
  institution  = {SNU Data Mining Center, Seoul National University},
  type         = {Technical Report},
  number       = {SNUDM-TR-2015-03},
  year         = {2015},
  url          = {https://dm.snu.ac.kr/static/docs/TR/SNUDM-TR-2015-03.pdf}
}

@INPROCEEDINGS{Lin2017,
  author={Lin, Lingli and Zhong, Shangping and Jia, Cunmin and Chen, Kaizhi},
  booktitle={2017 International Conference on Green Informatics (ICGI)}, 
  title={Insider Threat Detection Based on Deep Belief Network Feature Representation}, 
  year={2017},
  volume={},
  number={},
  pages={54-59},
  doi={10.1109/ICGI.2017.37}}

@Article{Saito2015,
  author    = {Saito, Takaya and Rehmsmeier, Marc},
  journal   = {PLOS ONE},
  title     = {The Precision-Recall Plot Is More Informative than the ROC Plot When Evaluating Binary Classifiers on Imbalanced Datasets},
  year      = {2015},
  issn      = {1932-6203},
  month     = mar,
  number    = {3},
  pages     = {e0118432},
  volume    = {10},
  doi       = {10.1371/journal.pone.0118432},
  editor    = {Brock, Guy},
  publisher = {Public Library of Science (PLoS)},
}

@InProceedings{Greitzer2013,
  author    = {Greitzer, Frank L. and Ferryman, Thomas A.},
  booktitle = {2013 IEEE Security and Privacy Workshops},
  title     = {Methods and Metrics for Evaluating Analytic Insider Threat Tools},
  year      = {2013},
  month     = may,
  pages     = {90--97},
  publisher = {IEEE},
  doi       = {10.1109/spw.2013.34},
}

@InProceedings{Davis2006,
  author     = {Davis, Jesse and Goadrich, Mark},
  booktitle  = {Proceedings of the 23rd international conference on Machine learning - ICML ’06},
  title      = {The relationship between Precision-Recall and ROC curves},
  year       = {2006},
  pages      = {233--240},
  publisher  = {ACM Press},
  series     = {ICML ’06},
  collection = {ICML ’06},
  doi        = {10.1145/1143844.1143874},
}

@Article{He2009,
  author    = {Haibo He and Garcia, E.A.},
  journal   = {IEEE Transactions on Knowledge and Data Engineering},
  title     = {Learning from Imbalanced Data},
  year      = {2009},
  issn      = {1041-4347},
  month     = sep,
  number    = {9},
  pages     = {1263--1284},
  volume    = {21},
  doi       = {10.1109/tkde.2008.239},
  publisher = {Institute of Electrical and Electronics Engineers (IEEE)},
}

@InProceedings{Glasser2013,
  author    = {Glasser, Joshua and Lindauer, Brian},
  booktitle = {2013 IEEE Security and Privacy Workshops},
  title     = {Bridging the Gap: A Pragmatic Approach to Generating Insider Threat Data},
  year      = {2013},
  month     = may,
  pages     = {98--104},
  publisher = {IEEE},
  doi       = {10.1109/spw.2013.37},
}

@Article{Le2021,
  author    = {Le, Duc C. and Zincir-Heywood, Nur},
  journal   = {IEEE Transactions on Network and Service Management},
  title     = {Anomaly Detection for Insider Threats Using Unsupervised Ensembles},
  year      = {2021},
  issn      = {2373-7379},
  month     = jun,
  number    = {2},
  pages     = {1152--1164},
  volume    = {18},
  doi       = {10.1109/tnsm.2021.3071928},
  publisher = {Institute of Electrical and Electronics Engineers (IEEE)},
}

@INPROCEEDINGS{Liu2018,
  author={Liu, Liu and De Vel, Olivier and Chen, Chao and Zhang, Jun and Xiang, Yang},
  booktitle={2018 IEEE International Conference on Data Mining Workshops (ICDMW)}, 
  title={Anomaly-Based Insider Threat Detection Using Deep Autoencoders}, 
  year={2018},
  volume={},
  number={},
  pages={39-48},
  doi={10.1109/ICDMW.2018.00014}}

@misc{Bengio2013Denoising,
      title={Generalized Denoising Auto-Encoders as Generative Models}, 
      author={Yoshua Bengio and Li Yao and Guillaume Alain and Pascal Vincent},
      year={2013},
      eprint={1305.6663},
      archivePrefix={arXiv},
      primaryClass={cs.LG},
      url={https://arxiv.org/abs/1305.6663}, 
}

@misc{Yuan2020,
      title={Deep Learning for Insider Threat Detection: Review, Challenges and Opportunities}, 
      author={Shuhan Yuan and Xintao Wu},
      year={2020},
      eprint={2005.12433},
      archivePrefix={arXiv},
      primaryClass={cs.CR},
      url={https://arxiv.org/abs/2005.12433}, 
}

@misc{che2016recurrentneuralnetworksmultivariate,
      title={Recurrent Neural Networks for Multivariate Time Series with Missing Values}, 
      author={Zhengping Che and Sanjay Purushotham and Kyunghyun Cho and David Sontag and Yan Liu},
      year={2016},
      eprint={1606.01865},
      archivePrefix={arXiv},
      primaryClass={cs.LG},
      url={https://arxiv.org/abs/1606.01865}, 
}

@misc{mei2017neuralhawkesprocessneurally,
      title={The Neural Hawkes Process: A Neurally Self-Modulating Multivariate Point Process}, 
      author={Hongyuan Mei and Jason Eisner},
      year={2017},
      eprint={1612.09328},
      archivePrefix={arXiv},
      primaryClass={cs.LG},
      url={https://arxiv.org/abs/1612.09328}, 
}

@misc{zaheer2018deepsets,
      title={Deep Sets}, 
      author={Manzil Zaheer and Satwik Kottur and Siamak Ravanbakhsh and Barnabas Poczos and Ruslan Salakhutdinov and Alexander Smola},
      year={2018},
      eprint={1703.06114},
      archivePrefix={arXiv},
      primaryClass={cs.LG},
      url={https://arxiv.org/abs/1703.06114}, 
}

\appendix
\section{Ray Tune Hyperparameters}
For applying experimental settings across scenario groupings, we utilized Ray, and applied various combinations of hyperparameters, which are  in table \ref{tab:search-space}. Ray applies these using the `tune.Tuner` class. 

For training, we utilize a g4dn.4xlarge instance in AWS, utilizing a single gpu for each experimental setting.

\section{Feature Space - CERT 5.2}
\begin{table}[ht]
\centering
\caption{\textbf{Ray Tune search-space hyperparameters.}
Sampled per trial by the ASHA scheduler.}
\label{tab:search-space}
\smallskip
\renewcommand{\arraystretch}{1.20}
\footnotesize
\begin{tabular}{@{} l l l @{}}
\toprule
\textbf{Parameter} & \textbf{Distribution} & \textbf{Range} \\
\midrule
\texttt{lr}
    & \texttt{loguniform}
    & $[10^{-4},\; 10^{-2}]$ \\
\texttt{latent\_dim}
    & \texttt{choice}
    & $\{32,\; 64,\; 96\}$ \\
\texttt{batch\_size}
    & \texttt{choice}
    & $\{32,\; 64\}$ \\
\texttt{dropout}
    & \texttt{choice}
    & $\{.05,\; .10,\; .15,\; .20,\; .25\}$ \\
\texttt{k\_neighbors}
    & \texttt{choice}
    & $\{2,\; 3,\; 4,\; 5,\; 6\}$ \\
\texttt{lambda\_t}
    & \texttt{uniform}
    & $[0.01,\; 0.07]$ \\
\texttt{lambda\_v}
    & \texttt{uniform}
    & $[0.01,\; 0.07]$ \\
\bottomrule
\end{tabular}
\renewcommand{\arraystretch}{1.0}
\normalsize
\end{table}

\begin{table}[ht]
\centering
\caption{\textbf{CERT r5.2 insider threat scenarios.}
Each scenario models a different insider threat archetype.
Attack windows are from ground-truth timestamps.}
\label{tab:cert52-scenarios}
\smallskip
\footnotesize
\renewcommand{\arraystretch}{1.25}
\begin{tabular}{@{} c c l @{}}
\toprule
\textbf{Scen.} & \textbf{Users} & \textbf{Attack Window} \\
\midrule
1 & 29 & Med.\ 6.0\,d (0.8\,h--9.0\,d) \\
2 & 30 & Med.\ 55.3\,d (43.2--59.2\,d) \\
3 & 10 & Med.\ 1.4\,d (1.4--1.6\,d)    \\
4 & 30 & Med.\ 74.8\,d (31.8--159.9\,d) \\
\midrule
\multicolumn{2}{c}{\textbf{Total: 99}}
  & Apr 2010--May 2011 \\
\bottomrule
\end{tabular}
\renewcommand{\arraystretch}{1.0}
\normalsize

\smallskip
\footnotesize
\renewcommand{\arraystretch}{1.15}
\begin{tabular}{@{} c p{5.8cm} @{}}
\toprule
\textbf{Scen.} & \textbf{Description} \\
\midrule
1 & Data exfiltration via USB + leak site. \\
2 & IP theft by departing employee. \\
3 & IT sabotage via keylogger. \\
4 & Credential misuse + email exfiltration. \\
\bottomrule
\end{tabular}
\renewcommand{\arraystretch}{1.0}
\normalsize
\end{table}

\end{document}